# Spectroscopic comprehension of Mott-Hubbard insulator to negative charge transfer metal transition in LaNi$_x$V$_{1-x}$O$_3$ thin films

Anupam Jana,[1] Sophia Sahoo,[1] Sourav Chowdhury,[1] Arup Kumar Mandal,[1] R. J. Choudhary,[1,*] D. M. Phase[1] and A. K. Raychaudhuri[2]

[1]UGC DAE Consortium for Scientific Research, Indore-452001

[2]CSIR- Central Glass and Ceramic Research Institute, Kolkata -700032

**Abstract:**

The room temperature (300 K) electronic structure of pulsed laser deposited LaNi$_x$V$_{1-x}$O$_3$ thin films have been demonstrated. The substitution of early-transition metal (TM) V in LaVO$_3$ thin films with late-TM Ni leads to the decreasing in out-of-plane lattice parameter. Doping of Ni does not alter the formal valence state of Ni and V in LaNi$_x$V$_{1-x}$O$_3$ thin films, divulging the absence of carrier doping into the system. The valence band spectrum is observed to comprise of incoherent structure owing to the localized V 3$d$ band along with the coherent structure at Fermi level. With increase in Ni concentration, the weight of the coherent feature increases, which divulges its origin to the Ni 3$d$-O 2$p$ hybridized band. The shift of Ni 3$d$-O 2$p$ hybridized band towards higher energy in Ni doped LaVO$_3$ films compared to the LaNiO$_3$ film endorses the modification in ligand to metal charge transfer (CT) energy. The Ni doping in Mott-Hubbard insulator LaVO$_3$ leads to the closure of Mott-Hubbard gap by building of spectral weight that provides the delocalized electrons for conduction. A transition from bandwidth control Mott-Hubbard insulator LaVO$_3$ to negative CT metallicity character in LaNiO$_3$ film is observed. The study reveals that unlike in Mott-Hubbard insulators where the strong Coulomb interaction between the 3$d$ electrons decides the electronic structure of the system, CT energy can deliver an additional degree of freedom to optimize material properties in Ni doped LaVO$_3$ films.

Corresponding Author

*E-mail: ram@csr.res.in

**Introduction:**

Since the discovery of colossal magnetoresistance (CMR) phenomenon in perovskite manganites, the high-Tc superconductors in cuprates, filling/bandwidth control metal to insulator transition (MIT) has escalated the research on the doped perovskite in TMOs systems [1-5]. In this aspect, $LaNi_xV_{1-x}O_3$ perovskite oxides would be very interesting because the parent compounds $LaVO_3$ (LVO) and $LaNiO_3$ (LNO) are electronically and electrically very different from each other. LVO displays the Mott-Hubbard type insulating character [6], while LNO shows the metallic nature at room temperature (RT) [7]. It can provide an interesting occurrence of insulator to metal transition as one increases $x$ in $LaNi_xV_{1-x}O_3$ and traverses a path between the end members of the series from paramagnetic Mott-Hubbard insulating LVO to Pauli paramagnetic metallic LNO. Besides, in $LaNi_xV_{1-x}O_3$, the $Ni^{3+}$ ($3d^7$) and $V^{3+}$ ($3d^2$) ions have different electron occupancies, which may reveal the competing spin and charge interactions and alter the various electron correlation energies leading to unusual physical properties.

The primitive cell of bulk LVO is orthorhombic with lattice parameters $a$ =5.55548 Å, $b$ = 7.84868 Å and $c$ = 5.55349 Å [8]. The strong electron correlation in LVO enforces the electron to be localized to their lattice site, consequently an insulating state emerges at RT and persists down to the low temperature (LT) [9,10]. LVO exhibits a structural transition from orthorhombic to monoclinic below 141 K and magnetic transition from paramagnetic to antiferromagnetic below 143 K [8,11]. Unlike iso-electronic $V_2O_3$ [4, 12], structural transition in LVO is not accompanied with MIT, however, conductivity is strongly suppressed in the monoclinic state of LVO due to the dispersion less nature of the conduction band [10,13]. On the other hand, the crystal structure of LNO is rhombohedral with Pauli paramagnetic behavior. Unlike other nickelates, LNO does not show a MIT and remains metallic down to 1.5 K [14]. Although, the $T^2$ dependence of resistivity at LT, heat capacity studies, and a Pauli paramagnetic-like susceptibility suggest an enhanced electron effective mass ($m^*$~10 $m_0$), which are attributed to strongly correlated Ni $3d$ $e_g$ electrons close to a MIT [15-17]. Therefore, in these compounds, various parameters such as electron-electron correlations, bandwidth, O $2p$-TM $3d$ hybridization strength, charge transfer energy, band occupancy etc. are some crucial microstructural parameters which tremendously influence the electrical, electronic, magnetic, specific heat and various other physical properties. [18-21]

Previous studies on doped perovskite TMOs, such as LaNi$_{1-x}$M$_x$O$_3$ ($M$ = Cr, Mn, Fe, or Co) compounds [22-25] suggested that there was a characteristic critical composition ($x_c$) which brings in the transition from the metallic to nonmetallic state in the series [26]. Interestingly, $x_c$ was found to be different for different TM elements, which indicated that the effects of electronic disorder also have crucial role on MIT [23]. Moreover, the electronic structure study in these doped systems also suggests that MIT is due to the potential mismatch between the substituent metal ion and Ni$^{+3}$ ion, which causes the transferring of hole states from near Fermi level (E$_F$) to an energy position above E$_F$ [24]. The substitution of TM with different ionic radius alters the TM-O-TM bond topology and leading to the modification in electronic bandwidth ($W$), which also has an important role on controlling MIT in doped systems [14,26-29]. Therefore, the mixed perovskite systems are the suitable ground to explore the MIT and its electronic manifestation to decipher the role of critical charge carrier densities, disorders as well as structural transition.

It should be noted here that these doping in LNO were mostly by the middle/late $M$ transition metals, with end members La$M$O$_3$ being mostly charge transfer insulators [30]. However, doping by early-TM element like V, with end member LaVO$_3$ being Mott-Hubbard insulator, is not studied. Therefore, in the present study we have investigated the electronic structure of LaNi$_x$V$_{1-x}$O$_3$ ($x$ =0, 0.2, 0.4, 0.5, 0.8 and 1.0) thin films at RT using photoemission and absorption spectroscopy techniques, where the substitution of V with Ni not only enhances the density of states (DOS) at the E$_F$, but also induces electronic disorder and the charge transfer between the two TM ions, which have huge implications of the microstructural parameters governing various physical properties.

**Experiment:**

Single phase LaNi$_x$V$_{1-x}$O$_3$ ($x$ = 0.0, 0.2, 0.4, 0.5, 0.8 and 1.0) thin films were grown on (001) oriented single-crystal LaAlO$_3$ (LAO) substrates using the pulsed laser deposition (PLD) technique. Fabrication of LVO film from single phase polycrystalline LaVO$_4$ was described elsewhere [6]. To grow the LNO film, we have followed the reference [31]. To grow the LaNi$_x$V$_{1-x}$O$_3$ films, number of dense pellets of each composition were used as a target for deposition. Bulk LaNi$_x$V$_{1-x}$O$_4$ were synthesized by a widely used solid-state reaction route. For the preparation of LaNi$_x$V$_{1-x}$O$_4$, stoichiometric amounts of highly pure (99.9%) La$_2$O$_3$, V$_2$O$_5$ and Ni$_2$O$_3$ were mixed and annealed at 600˚C for 6 hours (h) in air, followed by a second annealing at 1100˚C for 20 h, with an intermediate grinding for 12 hours. Lastly, the LaNi$_x$V$_{1-x}$O$_4$ targets were sintered at 1150°C for 15 h. Deposition of LaNi$_x$V$_{1-x}$O$_3$ films were carried out using single phase polycrystalline LaNi$_x$V$_{1-x}$O$_4$ target with a laser repetition rate of 3 Hz and a fluence of ≈ 1.8 J/cm$^2$. The target to substrate distance was maintained at 5.0 cm during deposition. The substrate temperature was kept at 650°C during deposition. To grow the LaNi$_x$V$_{1-x}$O$_3$ ($x$ = 0.0, 0.2, 0.4, 0.5) films, a vacuum near ≈5×10$^{-6}$ Torr were used, while for LaNi$_{0.8}$V$_{0.2}$O$_3$ film 70 mTorr oxygen partial pressure was used, as optimized for obtaining its single phase. After deposition, the LaNi$_x$V$_{1-x}$O$_3$ films were cooled to the RT at the rate of 10°C/ min in the same pressure as used during deposition. Thickness of all the grown films was about 20 to 25 nm.

The structural characterization of the grown films were done by a Bruker D2 Phaser x-ray diffractometer using Cu $K\alpha$ ($\lambda$ = 1.5406 Å) radiation. The composition of the grown films were verified by energy dispersive analysis of x-ray (EDAX) unit attached with field emission scanning electron microscope (FESEM) (FEI NOVA Nano Sem 450) and found to be consistent with the nominal composition. The x-ray photoemission spectroscopy (XPS) measurements of the grown films were carried out with Al $K\alpha$ (1486.6 eV) x-ray source using the Omicron energy analyzer (EA-125, Germany). RT valence band spectra (VBS) of films were recorded at photon energy of 52 eV at the angle integrated photo emission spectroscopy (AIPES) beamline on Indus-1 synchrotron source at RRCAT, Indore, India. The base pressure in the experimental chamber during measurements was of the order of 10$^{-10}$ Torr. Prior to the photoemission measurements, the surface of all the films were cleaned *in situ* using low energy Ar$^+$ ions. For calibration of binding energies, Au foil was kept in electrical contact with the sample and the Fermi level (E$_F$) was aligned using the VBS of Au foil. The total instrumental resolution was about 300 meV at $h\nu$ = 52 eV. To investigate the unoccupied states of the grown films, x-ray absorption near edge spectroscopy (XANES) was carried out at RT in the total electron yield (TEY) mode at the beam line BL-01, Indus-2 synchrotron source at RRCAT, Indore, India. The energy resolution during XANES measurements across the measured energy range was estimated to be ~250 meV.

## Results and Discussions

**Crystal Structure:**

Figure 1(a), shows the RT θ-2θ x-ray diffraction (XRD) patterns of LaNi$_x$V$_{1-x}$O$_3$ (x = 0.0, 0.2, 0.4, 0.5, 0.8 and 1.0) films along with single-crystalline LAO substrate. The diffraction pattern reveals

the single phase growth of all the films along (00$l$) direction. The calculated pseudocubic out-of-plane lattice parameter of grown LVO ($x = 0.0$) film is found to be 4.04 Å. The pseudocubic lattice parameter of bulk LVO is 3.93 Å [32]. The lattice mismatch with the underlying LAO substrate (lattice parameter of 3.78 Å) will induce in-plane compressive strain in the film leading to the observed enhanced out-of-plane lattice parameter of LVO film with respect to its bulk counterpart. On increasing Ni concentration, the shift in the (002) peak towards higher angle confirms that the out-of-plane lattice parameter of the grown films gradually decreases, as shown in the Fig. 1(b). As the $Ni^{3+}$ has an ionic radius of 0.74 Å in an octahedron environment, whereas that of $V^{3+}$ is 0.78 Å, a decrease in the lattice parameter and unit cell volume is expected on increasing the Ni content in LVO. Thus, the pseudocubic out-of-plane lattice parameter of $LaNi_xV_{1-x}O_3$ films is found to decrease from 3.98 Å in $LaNi_{0.2}V_{0.8}O_3$ to 3.88 Å in $LaNi_{0.8}V_{0.2}O_3$. The obtained out-of-plane lattice parameter 3.84 Å of LNO ($LaNi_xV_{1-x}O_3$; $x = 1.0$) film, which is slightly elongated compared to the pseudocubic bulk LNO (3.830 Å), is comparable to the previous results [31]. Usually the pseudocubic lattice constant of doped compounds is linearly dependent on doping level [33]. As the pseudocubic lattice parameter of bulk LNO is $a_p^{LNO}= 3.83$ Å and bulk LVO is $a_p^{LVO}= 3.93$ Å, hence the pseudocubic lattice parameter for doped compounds $LaNi_xV_{1-x}O_3$ should vary linearly following the equation $a_p^{LNi_xV_{1-x}O_3} = a_p^{LNO}(x) + a_p^{LVO}(1-x)$. However, as shown in the Fig. 1(b), the obtained out-of plane lattice parameter of $LaNi_xV_{1-x}O_3$ films are elongated as compared to the expected pseudo-cubic bulk $LaNi_xV_{1-x}O_3$, which arises because the underlying LAO substrate will induce in-plane compressive strain in the film owing to psuedomorphic growth.

**Electronic structure:**

**Core level photoelectron spectrum:**

LNO is a paramagnetic metal, so Ni doping in LVO can facilitate electron delocalization. Such modification will also be manifested in the core level spectra via final states of core-level photoemission. Figure 2(a) depicts the V $2p$ and O $1s$ core level XPS of LVO and the intermediate compositional $LaNi_{0.5}V_{0.5}O_3$ films. The spectrum consists of spin-orbit split V $2p$ states along with O $1s$ state. To estimate the position of individual features of LVO, we have fitted the spectrum with a combination of Gaussian and Lorentzian functions and the background of the spectrum was corrected by Shirley function as describe elsewhere [6]. The obtained binding energy (BE) position of the spin-orbit split V $2p_{3/2}$ and V $2p_{1/2}$ states for LVO are 515.6 eV and 522.8 eV, in conformity with the $V^{3+}$ state in LVO bulk [34]. Similar BE positions of the V $2p_{3/2}$ and V $2p_{1/2}$ states are also observed for $LaNi_{0.5}V_{0.5}O_3$ film, confirming the formal charge state of V to remain in 3+ after Ni doping in LVO. There is an overall broadening of spectral feature of V $2p$ on substituting Ni in LVO. Generally, the asymmetry and broadening in the V $2p$ photoelectron spectrum of $V^{3+}$ state arise due to the occurrence of multiplet structure owing to the presence of unpaired electrons [35]. However, a slight enhanced broadening of V $2p$ photoelectron spectrum is evident in $LaNi_{0.5}V_{0.5}O_3$, in the form of an appearance of a shoulder structure at lower BE to the main V $2p_{3/2}$ peak as shown clearly in Fig. 2(b). A similar behavior was reported in the V $2p$ photoelectron spectra of $V_2O_3$ and $VO_2$ observed across the MIT [36,37]. Such shoulder structure in V $2p_{3/2}$ peak of $LaNi_{0.5}V_{0.5}O_3$ film arises due to the extra screening of V $2p$ electrons from coherent states, created at the $E_F$ upon Ni doping in LVO as shown schematically in Fig. 2(c). The O $1s$ core-level photoelectron spectra show a clear spectral shape change; a narrow symmetric peak in insulating LVO to a broad asymmetric line shape in $LaNi_{0.5}V_{0.5}O_3$ as shown in the Fig. 2(d). The asymmetric to symmetric line shape changes are similar to those observed in the O $1s$ spectra of $V_2O_3$ and $VO_2$ across MIT [36,37] and metallic $SrVO_3$, $CaVO_3$ compounds [38]. Thus, a change in line shape of O $1s$ spectrum in $LaNi_{0.5}V_{0.5}O_3$ film indicates the creation of coherent states at $E_F$, is a signature of metallic nature of $LaNi_{0.5}V_{0.5}O_3$ film. The O $1s$ spectrum of LNO film also shows narrow line shape as shown in the Fig. 2(d). Besides, the existence of different covalent Ni-O and V-O bond environment in the $LaNi_{0.5}V_{0.5}O_3$ film could also be the reason for broad O $1s$ line shape compared to the LVO and LNO films.

We have further simulated the V $2p$ core-level spectra using the charge transfer multiplet calculation [39], which very much represents the experimental spectrum as shown in the Fig. 2(a). The simulated spectrum was convoluted with a 0.7 eV Gaussian to account for the instrumental resolution. Apart from the sharp V $2p_{3/2}$ and V $2p_{1/2}$ main peaks at 515.6 eV and 523.4 eV, a weak intense satellite structure is also observed around 532.5 eV BE position. The cluster model calculation on LVO suggests that the main peaks V $2p_{3/2}$ and V $2p_{1/2}$ states correspond to the well screened $\underline{c}3d^3\underline{L}$ ($\underline{c}$: hole in the core level, $\underline{L}$: hole in O $2p$ band) final state configuration, whereas the CT satellite structure corresponds to the poorly screened $\underline{c}3d^2$ configurations and appears at the higher BE side of the O $1s$ peak (~532 eV) and around 542 eV respectively [38]. As, the satellite structure of V $2p_{3/2}$ state appears at 532.5 eV and is buried in the higher BE side of the O $1s$ peak, it causes a slight asymmetric shape of O $1s$ even in insulating LVO.

**Near edge x-ray absorption spectroscopy (NEXAS):**

To understand the modifications in local electronic structure of Ni doped LVO films, we performed the near edge x-ray absorption spectroscopy (NEXAS) of LaNi$_x$V$_{1-x}$O$_3$ films. Figure 3(a) shows the V $L_{3,2}$-edge absorption spectra of the LaNi$_x$V$_{1-x}$O$_3$ films measured at RT. The corresponding V $L$-edges spectra are further compared with the spectrum of trivalent V$_2$O$_3$ (V$^{3+}$: $3d^2$) and pentavalent V$_2$O$_5$ (V$^{5+}$: $3d^0$) compounds. The energy position of the absorption maxima of $L_3$ edge of LaNi$_x$V$_{1-x}$O$_3$ films is similar to those of V$_2$O$_3$ but quite different from those of V$_2$O$_5$ and VO$_2$ [40], which confirms that V ions are formally in trivalent (V$^{3+}$:$3d^2$) state in the present studied LaNi$_x$V$_{1-x}$O$_3$ films. The maxima of V $L_{3,2}$ absorption peaks of the LaNi$_x$V$_{1-x}$O$_3$ films appear at the position of 517.8 eV and 523.8 eV, in conformity with the previous XAS measurements of single crystal and thin film of LVO and other RVO$_3$ compounds [6,41, 42]. However, substantial change is observed in the line shape of the $L_3$ absorption peak. As the doping of Ni increases, the FWHM of the V $L_3$ peak decreases, resulting in the narrowing of the $L_3$ peak and concomitantly the more pronounced low energy shoulder is observed with increasing $x$ in LaNi$_x$V$_{1-x}$O$_3$ films. This trend is shown more clearly at the top of the Fig. 3(c), where two XAS of $x$=0.8 (LaNi$_{0.8}$V$_{0.2}$O$_3$) and $x$=0 (LaVO$_3$) are normalized at the maximum intensity. Interestingly, this shoulder structure shifts towards lower energy with increasing Ni concentration in LVO. Such enhancement of shoulder structure was also observed in Mn $L_3$ edge of SrMn$_{1-x}$Fe$_x$O$_3$ with increase in $x$ [43] and was attributed to the charge transfer effect between the Mn $3d$ and O $2p$ orbitals. Generally, the line shape of the V $L_{3,2}$ absorption peaks depend on the atomic multiplet states, crystal field and ligand to metal CT effect [44]. The crystal field splitting is caused by the interaction of metal $d$ electrons with the surroundings ligand ions in the crystals, whereas, the multiplet states are originated due to the intra-atomic $3d$-$3d$ Coulomb and $2p$-$3d$ Coulomb and exchange interactions [44]. Moreover, these effects enormously depend on the local site geometry in a crystal, such as M-O-M networks and the corresponding hybridization between metal $3d$ and oxygen $2p$ ions [49]. Thus, the more pronounced shoulder structure of the V $L_3$ peak and its shifting towards lower energy with increasing Ni doping in LVO imply the modulation in local structural symmetry and corresponding energetics. To understand the mechanism behind the emergence of pre-peak structure in V $L_3$ edge with Ni doping, we have further simulated the V$^{3+}$ $L$-edge using charge transfer multiplet calculation [39].

Figure 3(b) shows the O $K$-edge NEXAS of all LaNi$_x$V$_{1-x}$O$_3$ films. The O $K$-edge spectrum corresponds to the transition of an electron from the O 1$s$ core-level to the unoccupied O 2$p$ states hybridized with the metal 3$d$ and rare-earth 5$d$ states [46]. For LVO (V$^{3+}$: 3$d^2$), the lower energy structure in the photon energy range of 529.0 eV to 532.5 eV primarily corresponds to the O 2$p$ character mixed with the unoccupied part of the V 3$d$ state, whereas other broad structures appearing around 535.0 eV and 540-545 eV are due to the O 2$p$ character mixed with the La 5$d$ and V 4$s$-4$p$ hybridized states [6] as shown in Fig. 3(b). In LVO, perovskite cubic octahedral ($O_h$) symmetry splits the V 3$d$ levels into three fold $t_{2g}$ and two fold $e_g$ levels. Thus, the broad feature appearing in the photon energy range of 529.0 eV to 532.5 eV is attributed to the hybridized O 2$p$ -V $t_{2g}$ and $e_g$ states respectively. With respect to the thicker and single crystal LVO studies, in the studied thin film of LVO these peaks appear closer and form a broad structure, revealing lesser effective crystal field energy ($\Delta_{CF}$) (~ 1 eV) than 1.3 eV and 1.8 eV obtained for thicker and single crystal LVO respectively [6,38]. Such reduction in $\Delta_{CF}$ value in the studied LVO film arises due to the modification in V-O bond length, because the $\Delta_{CF}$ is very sensitive to changes of the interatomic distances V-O since $\Delta_{CF} \propto$ (V-O)$^{-5}$ [47].

In LaNi$_x$V$_{1-x}$O$_3$ ($x$ = 0.2, 0.4, 0.5, 0.8 and 1.0) films, a sharp feature appears at the photon energy of 529.2 eV and remains fixed in energy up to 50% Ni substitution in LVO as shown in the Fig. 3(b). Interestingly this feature shifts further towards lower photon energy with the enhanced Ni concentration and finally appears at 528.2 eV for LNO ($x$ = 1.0) similar to previous reports [48,49]. This peak is attributed to the unoccupied Ni 3$d$ states, mostly dominated by $e_g$ character. In rare-earth nickelates, due to the strong Ni 3$d$-O 2$p$ hybridization, the ground state is composed of a mixture of configuration 3$d^7$, 3$d^8\underline{L}$, 3$d^9\underline{L}^2$ ($\underline{L}$: is a ligand hole), instead of purely ionic configuration (Ni$^{3+}$O$^{2-}$). Resultantly, the ground state has an appreciable amount of 3$d^8\underline{L}$ character that causes such pre-edge structure, which is attributed to the transition 3$d^8\underline{L} \rightarrow \underline{c}3d^8$ ($\underline{c}$: O 1$s$ core-hole) [50] and is treated as a mark of Ni 3$d$-O 2$p$ hybridization strength. It is important to mention here that with reduced Ni concentration, obtained pre-edge feature in O $K$-edge of LaNi$_x$V$_{1-x}$O$_3$ films is shifted towards higher photon energy by around 1.0 eV as compared to LNO (Ni$^{3+}$) film. Earlier the movement of such pre-peak towards higher energy was reported to occur in epitaxial LNO films with enhanced substrate induced strain [19] and was ascribed to the reduced Ni-O bond covalency and corresponding enhancement of charge transfer energy ($\Delta$) [51]. Therefore, such shifting of pre-peak towards higher energy in the present studied LaNi$_x$V$_{1-x}$O$_3$ films could be explained in terms of the reduced covalency of Ni, as expected from the enhanced Ni-O distances with reduction in Ni concentration in LVO. Similar correlation between Cu-O distance and the energy threshold of the pre-peak has been drawn for the insulating precursors of several high T$_C$-superconductors [52]. Although, similar pre-peak was also present in O 1$s$ XAS of NiO (Ni$^{2+}$) corresponding to $3d^9\underline{L} \rightarrow \underline{c}3d^9$ transition [49], however its energy position (532.0 eV) is about 3.0 eV higher than the pre-peak of LaNi$_{0.2}$V$_{0.8}$O$_3$ film that invalidates the presence of Ni$^{2+}$ in the present studied LaNi$_x$V$_{1-x}$O$_3$ films. Furthermore, splitting in this pre-peak was observed in the oxygen deficient LNO compounds as shown in a study by Abbate et al [53]. Based on cluster model calculation, splitting has been attributed to the presence of two non-equivalent crystallographic sites owing to the oxygen deficiency. However, no such splitting has been observed in the pre-peak of presently studied LaNi$_x$V$_{1-x}$O$_3$ films, which further rules out the presence of oxygen defects states.

Importantly, Ni doping in LVO leads to the substantial modification in the energy position of unoccupied V 3$d$ state as shown in the Fig. 3(b). A peak observed at 532.4 eV in the O $K$-edge of LaNi$_{0.2}$V$_{0.8}$O$_3$ film, which is absent in LNO film, represents the unoccupied V 3$d$ states. It appears that with Ni doping in LVO films, a spectral feature representing the Ni 3$d$-O 2$p$ hybridization emerges at lower photon energy side of the unoccupied V 3$d$ band, which subsequently pushes the V 3$d$ band towards higher energy. Moreover, the V $t_{2g}$ and $e_g$ states, which are merged together and form a broad structure in LVO, appear to be separated with two isolated relatively narrow features in Ni doped LVO films as shown in the Fig. 3(b). The separation between the V $t_{2g}$ and $e_g$ states increases, suggesting an increase in the crystal filed splitting with Ni doping in LVO. The spectral feature at 535.4 eV and broad structures appearing around 540-545 eV for all the LaNi$_x$V$_{1-x}$O$_3$ films correspond to the unoccupied La 5$d$ and Ni/V 4$s$-4$p$ hybridized states [48].

In the V $L$-edge spectra, it should be noted that on going from LaVO$_3$ to LaNi$_{0.8}$V$_{0.2}$O$_3$, the V $L_{3,2}$ absorption peak narrows and shoulder or pre-peak feature is enhanced. To better understand this trend, we compare the V 2$p$ XAS spectra of LaVO$_3$ ($x$=0) and LaNi$_{0.8}$V$_{0.2}$O$_3$ ($x$= 0.8) with those of simulated V$^{3+}$ $L$-edge spectrum using charge transfer multiplet program for x-ray absorption spectroscopy (CTM4XAS) [39] as shown in Fig. 3(c). The shape of V $L_3$ and $L_2$ lines is very sensitive to the $\Delta_{CF}$ along with the V$^{3+}$ ground state ($2p^63d^2$) and excited state ($2p^53d^3$) multiplets, which can be controlled by the two-particle interaction parameter. Therefore, we performed calculations by varying the reduction of Slater integrals, the crystal field splitting ($\Delta_{CF}$), charge transfer energy ($\Delta$), $d$-$d$ interaction energy ($U$), and O $2p$-V $3d$ hybridization strength. Slater integrals were reduced to 70% of their atomic values to simulate both spectra. The simulated $L_{3,2}$ edge spectrum of $V^{3+}$ with $d^2$ ($^3T_1$) ground state along with the crystal field in $O_h$ symmetry of 1.0 eV, $\Delta = 6.8$ eV, $U_{2p3d} - U_{3d3d} = 2.2$ eV, and hopping parameter $V(e_g) = 2.8$ eV matches well with the experimentally observed V $L$-edge spectrum of LaVO$_3$ (marked as feature **S** in the Fig. 3(c)). The simulated line spectrum is convoluted with Lorentzian line shape equal to the lifetime width 0.4 eV of the $L_3$ core-hole and with a Gaussian line width 0.25 eV to represent instrumental broadening. Although both the simulated profiles agree well with the experimental profiles of V $L$-edge, the minor source of discrepancy could be due to the non-uniformity of the core-hole life time broadening. To understand the modifications in V $L$-edge spectrum with Ni doping, we further alter the different parameters to simulate the $V^{3+}$ $L$-edge spectrum for LaNi$_{0.8}$V$_{0.2}$O$_3$ film. It is observed that with higher $\Delta_{CF} = 2.4$ eV and lower $\Delta$ value (~3.0 eV), a well intense pre-peak generated. It is known that the crystal field interaction mixes states with different $L$ values in the ground state such that the transition to other final states is allowed with the same spin [54]. Such mixing gives rise to the broadening of the peaks and a gradual appearance of new peaks with increasing crystal fields. Our observed intense pre-peak in LaNi$_{0.8}$V$_{0.2}$O$_3$ film and well simulated $V^{3+}$ $L$-edge spectrum marked as **S\*** in Fig. 3(c) with considering higher $\Delta_{CF}$ and lower $\Delta$ compare to LVO film suggests an increase in metal-ligand hybridization with Ni doping in LVO. The enhanced $\Delta_{CF}$ in LaNi$_x$V$_{1-x}$O$_3$ films with higher $x$ values is further confirmed in O $K$-edge oxygen spectra of Ni doped films. Although, previously enhanced charge transfer energy or weaker covalency [43] was suggested to be the primary cause for the broadening of $L_{3,2}$ absorption peaks, however our observation from simulation suggests that peak broadening is very much sensitive to the variation of $\Delta_{CF}$.

Figure 4(a) shows the La $M_{5,4}$ and Ni $L_3$-edge spectra of LaNi$_x$V$_{1-x}$O$_3$ films measured at 300 K. The La $M_{5,4}$ absorption peaks arise due to transition of an electron from spin-orbit split La $3d$ core level to the unoccupied La $4f$ states. Observed energy positions of La $M_{5,4}$-edge for all the films match well with that of La$^{3+}$ reported in literature [55], confirming the 3+ charge state of La. Owing to very small energy separation between La $3d_{3/2}$ and Ni $2p_{3/2}$ core levels, the La $M_4$ and Ni $L_3$-edges overlap as shown in the Fig. 4(a). [56]. Hence, in order to study the Ni $L$-edge spectra, it was extracted by fitting multiple Lorentzian peaks to the overlapped region and then subtracting La $M_4$- edge from raw data as shown in the Fig. 4(b) [57]. The extracted Ni $L_{32}$ edge spectra of LaNi$_x$V$_{1-x}$O$_3$ films are shown in Fig. 4(c). Interestingly, like V $L$-edge, the Ni $L$-edge also shows drastic variation with increasing Ni concentration in LVO. The observed $L_3$-edge for higher Ni concentrated LaNi$_{0.8}$V$_{0.2}$O$_3$ and LNO films show a single peak structure, while lower Ni concentrated LaNi$_{0.5}$V$_{0.5}$O$_3$, LaNi$_{0.4}$V$_{0.6}$O$_3$ films show double peak structure. Moreover, the $L_3$-edge line shape of LaNi$_{0.8}$V$_{0.2}$O$_3$ film is much narrower compared to the LNO film. The Ni $L$-edge spectrum of nickelates exhibits distinct double-peak structures originating from $2p^63d^8$–$2p^53d^9$ and $2p^63d^8\underline{L}^n$–$2p^53d^9\underline{L}^n$ multiplet transitions [49]. However, the relative intensity and the separation between the double-peak depends on the energetic balance among microscopic parameters, including site energy, charge-transfer energy and rare-earth Ni hybridization energy, which are expected to vary in LaNi$_x$V$_{1-x}$O$_3$ films [49]. The energy position of Ni $L_3$ edge confirms the Ni$^{3+}$ valence state in Ni doped LVO films. The trivalent valence states for both V and Ni ions in LaNi$_x$V$_{1-x}$O$_3$ films suggest that Ni doping in LVO does not lead to the carrier doping in the valence band of LVO, rather the modification in bandwidth due to alteration in TM-oxygen hybridization upon Ni doping.

**Valence band spectra:**

The VBS of LaNi$_x$V$_{1-x}$O$_3$ ($x$=0.0, 0.5, 0.8 and 1.0) films are recorded at photon energy of 56 eV at 300 K is shown in Fig. 5(a). It has been shown previously [6,58] that for LVO, the spectrum in the BE range between the 0.5 eV to 3.0 eV is due to the dominant V $3d$ character and the energy region between 3.0 to 9.0 eV is mostly dominated by the O $2p$ contribution. The RPES study of LVO film at 300 K [6] further confirmed the dominated V $3d$ states with $3d^{n-1}$ final state character of the spectral feature centered at 1.5 eV and attributed to lower Hubbard band (LHB), while the feature around 6.9 eV represented V $3d$ – O $2p$ hybridized bond with $3d^n\underline{L}$ final state character schematically shown in the Fig. 5(c). These observations are also corroborated with the theoretically calculated density of states (DOS) [13]. For LNO, previous studies suggest that BE region 7.5 to 3.5 eV is attributed to the O $2p$ bonding states and the region between 3.5 to 1.5 eV is assigned to the O $2p$ non-bonding states, while the spectrum from 1.5 to E$_F$ level corresponds to the Ni $3d$ states [49]. RPES study of LaNiO$_{3-\delta}$ thin films also suggest that the features near E$_F$ have dominant Ni $3d$ band [59], however substantial amount O $2p$ states in the form of hybridization are also present. The features at about 6.0 eV is due to the admixture of La $4d$ and O $2p$ states. Considering the electron configuration $t_{2g}^6 e_g^1$ for Ni$^{3+}$ in LNO, it appears that the weak intense feature at E$_F$ is related to the $e_g$ band while the feature at 1.0 eV is associated with the $t_{2g}$ band [18], schematically shown in the Fig. 5(c). The presence (absence) of finite spectral DOS at the $E_F$ confirms the metallic (insulating) nature of LNO (LVO) film. The overall spectral features of VB of LaNi$_{0.5}$V$_{0.5}$O$_3$ film (Fig. 5(a)) appear somewhat similar to the VB of LVO; however, strong modification is observed at around 1.5 eV BE near the $E_F$, which turned to be more asymmetric after Ni doping. To understand more on this aspect, we further normalized the spectra of LaNi$_{0.5}$V$_{0.5}$O$_3$ and LVO film at 1.5 eV, as shown in the Fig. 5(b). Beside intense 1.5 eV feature, an additional shoulder structure appears at 0.3 eV in LaNi$_{0.5}$V$_{0.5}$O$_3$ film and the tail of this shoulder structure crosses the $E_F$, resulting in an enhancement of spectral DOS at $E_F$ of LaNi$_{0.5}$V$_{0.5}$O$_3$. Such spectral enhancement at $E_F$ represents the coherent states, which emerges due to the delocalized conduction electrons in LaNi$_{0.5}$V$_{0.5}$O$_3$ film. Such spectral DOS at $E_F$ further enhances with the Ni doping in LVO as seen in the VBS of LaNi$_{0.8}$V$_{0.2}$O$_3$ film. In LNO, the spectral feature at the $E_F$ is strongly dominated by the occupied Ni $3d$ states hybridized with O $2p$ bands. Thus, the observed shoulder feature near across $E_F$ of LaNi$_x$V$_{1-x}$O$_3$ ($x$ = 0.5 and 0.8) films is possibly due to the presence of the occupied Ni $3d$ states, schematically shown in the Fig. 5(c), which provides delocalized electrons responsible for metallic conduction in the system.

To further understand the atomic origin of the spectral features, primarily near $E_F$ in VBS of LaNi$_x$V$_{1-x}$O$_3$ films, we performed the V $3p \rightarrow 3d$ RPES of LaNi$_{0.5}$V$_{0.5}$O$_3$ film at 300 K and compared with LVO film. The on and off-resonance spectra are recorded at the incident radiation of 52 eV and 40 eV respectively, as shown in the Fig. 6(a). Comprehensive study on the RPES of LVO film at 300 K is described elsewhere [6]. The difference spectrum, obtained by subtracting the off-resonance spectrum and inelastic background from the on-resonance spectrum, represents roughly the partial spectral weight distribution of the V $3d$ electrons. So we consider this difference spectrum as V $3d$ spectral DOS of LVO film. It is important to note that the extracted V $3d$ spectral DOS of LVO film matches well with the calculated V $3d$ partial-DOS (PDOS) of strained orthorhombic LVO obtained from the density functional theory (DFT) using generalized gradient approximation (GGA) for exchange correlation as shown in the Fig. 6(c) [13]. It is evident for LaNi$_{0.5}$V$_{0.5}$O$_3$ film, shown in the Fig. 6(b), the spectral feature at 1.5 eV is strongly enhanced with V$3p \rightarrow 3d$ excitation threshold akin to the LVO film, indicating that feature at 1.5 eV has strong V $3d$ character. We have followed the same extraction procedure as described above to determine the V $3d$ spectral DOS of LaNi$_{0.5}$V$_{0.5}$O$_3$ film. Figure 6(c) shows the extracted V $3d$ spectral DOS of LVO and LaNi$_{0.5}$V$_{0.5}$O$_3$ films. It is found that V 3$d$ spectral DOS exhibits a double peak structure centered at 1.5 eV and around 6.9 eV BE respectively for both LVO and LaNi$_{0.5}$V$_{0.5}$O$_3$ films. The sharp peak appears at ~1.5 eV below E$_F$ in the V 3$d$ spectral DOS of LVO and LaNi$_{0.5}$V$_{0.5}$O$_3$ films, and is attributed to the dominant V $t_{2g}$ band [6,58]. Whereas, the broad feature around 6.0 -7.0 eV BE in V 3$d$ spectral DOS represents the O 2$p$-V 3$d$ hybridized states [6,58,13]. Interestingly, the V 3$d$ spectral DOS around O 2$p$-V 3$d$ hybridized region appears extended in LaNi$_{0.5}$V$_{0.5}$O$_3$ compared to the LVO film, indicating increase in V 3$d$- O 2$p$ hybridization with Ni doping in LVO film. Moreover, the spectral intensity ratio $I$ (1.5 eV)/$I$ (6.9 eV) decreases, suggesting a transfer of V 3$d$ spectral weight from localized incoherent state to V 3$d$ - O 2$p$ hybridized state with Ni doping in LVO as evident in Fig. 6(c). It is important to highlight here that the extracted V 3$d$ spectral DOS of LaNi$_{0.5}$V$_{0.5}$O$_3$ film does not display spectral intensity at E$_F$ [$I$ (E$_F$)], which suggests that V 3$d$ electrons are strongly localized even with Ni doping in LVO. So, the shoulder feature appearing near E$_F$ of the VB of LaNi$_{0.5}$V$_{0.5}$O$_3$ film, which is also responsible for finite spectral DOS at E$_F$, could be attributed to the Ni 3$d$ states. To confirm this we have also recorded the VBS of LNO film at on and off-resonance with incident radiation of 64 eV and 40 eV respectively shown in Fig. 6(d). Difference spectrum represents the spectral weight distribution of Ni 3$d$ electron, which reveals a finite Ni 3$d$ contribution near at E$_F$, as clearly shown in the inset of Fig. 6(d). Thus, the shoulder feature in LaNi$_{0.5}$V$_{0.5}$O$_3$ confirms that the delocalized Ni 3$d$ electrons are responsible for the metallic conduction in this system akin to the LNO [49,18]. Therefore, in LaNi$_x$V$_{1-x}$O$_3$ films, besides the presence of incoherent structure which represents the localized V 3$d$ band, a coherent structure also appears which represents the Ni 3$d$ band near and at E$_F$.

**Electronic band structure across the Fermi level (E$_F$):**

To understand the modulation of low energy charge fluctuation across the E$_F$ and the evolution of electronic states near E$_F$ with the Ni substitution in LVO, the experimental valence band (VB) and conduction band (CB) of LaNi$_x$V$_{1-x}$O$_3$ ($x$=0.0, 0.5 and 1.0) films are combined together, as shown in Fig. 7(a). For the CB, oxygen $K$-edge spectrum has been used, as it can be considered to represent the most weighted unoccupied character of TM 3$d$, TM 4$sp$ and La 5$d$ sates via the hybridization with O 2$p$ states. In addition, the photo induced core-hole effect on the final state DOS is less severe compared to the TM 2$p$ edge [46]. To plot the CB, oxygen $K$-edge of the LaNi$_x$V$_{1-x}$O$_3$ films was subtracted from the BE position of the rising tail of the oxygen core level photoelectron spectrum [60,61]. The electronic structure of LaNi$_x$V$_{1-x}$O$_3$ films obtained from the experimental occupied and unoccupied DOS indicates a substantial modification in spectral DOS, mostly appearing near across the E$_F$ with Ni doping in LVO film. In LVO film, as shown in the Fig. 7(a), the first spectral feature in the VB appearing at −1.5 eV below the E$_F$ has the dominant occupied V 3$d$ character as confirmed through the RPES study [6] and DFT calculation [13]. Thus, this structure, normally termed as an incoherent feature in one-electron-removal spectrum, is ascribed as a spectral signature of the lower Hubbard band (LHB) [6,58]. In the CB, first prominent broad structure is observed from +2.1 to +3.0 eV above the E$_F$. The DFT calculation and the bremsstrahlung isochromat (BI) spectroscopic measurements on bulk LVO [13,58] confirmed that such spectral structure appeared due to the unoccupied V 3$d$ $t_{2g}$ and $e_g$ structure. Hence, the unoccupied V $t_{2g}$ state appearing at ~ +2.1 eV above E$_F$ is attributed to the spectral signature of upper Hubbard band (UHB). The energy separation between the LHB and UHB was found to be ~3.6 eV, which is the energy difference between the $d^{n-1}$ and $d^{n+1}$ electronic state and related to the electron correlation strength ($U$) of the LVO film, clearly shown in the inset of Fig. 7(a). Such high electron correlation energy compared to the kinetic energy (bandwidth) of the LVO at 300 K impedes the electron delocalization, resultantly an insulating state is emerged through electron localization at their lattice site even in partially occupied V $t_{2g}^2$ system [4,5]. The calculated band gap is about 1.0 eV as obtained from the combined spectra. The detailed analysis of the VBS describe elsewhere [6] confirms that the O 2$p$ band lies well below the V 3$d$ band in LVO film, as schematically shown in the Fig. 7(b), divulging the Mott-Hubbard type insulating nature of LVO film at RT [6].

Combined spectra shows a large spectral DOS at the E$_F$ of LNO film confirming its metallic nature at RT shown in Fig. 7(a) [49]. The VBS of LNO mimics the band structure of a negative charge transfer metals (-Δ), as has been predicted earlier [62,63]. In these negative Δ materials, the band gap does not belong to either a $p$-$d$ type or a $d$-$d$ type, rather it is a $p$-$p$ type and the charge fluctuations can be viewed as $d^{n+1}\underline{L}$ $d^{n+1}\underline{L}$ = $d^{n+1}\underline{L}^2$ $d^{n+1}$ [61,62]. Thus, a considerable amount O 2$p$ states are presents in the form of hybridization with the Ni 3$d$ states at the E$_F$, schematically shown in the Fig. 7(b), which effectively reduces the energy cost of transferring electron from O 2$p$ sates to unoccupied Ni 3$d$-O 2$p$ hybridized sate. Resultantly, the effective value of Δ decreases and hence O 2$p$-hole along with Ni 3$d$ electrons are participating in low energy conduction process [18].

In LaNi$_{0.5}$V$_{0.5}$O$_3$ film, besides the presence of the incoherent structure at −1.5 eV, corresponding to the correlated localized V 3$d$ electronic states, a coherent structure also appears at the lower energy side of V 3$d$ band corresponding to the hybridized Ni 3$d$-O 2$p$ band. In the CB, the first feature centered at +0.5 eV is also attributed to the Ni 3$d$-O 2$p$ hybridized band. The tail of this band is extended to the E$_F$ and consequently manifests a finite DOS at E$_F$ as shown clearly in Fig. 7(a). Therefore, with the Ni doping in LVO, Ni 3$d$-O 2$p$ hybridized states are emerging near across E$_F$. Such enhanced spectral DOS at E$_F$ is an outcome of the enhanced bandwidth of the Ni 3$d$-O 2$p$ hybridized band with Ni doping, which provides delocalized conduction electrons [19]. A sharp feature appearing at the +3.2 eV in CB of LaNi$_{0.5}$V$_{0.5}$O$_3$ film is due to the unoccupied V 3$d$ band. It is important to note here that in LaNi$_{0.5}$V$_{0.5}$O$_3$ film the energy separation between occupied and unoccupied V 3$d$ states, i.e. the energy difference between V 3$d^1$ and V 3$d^3$ electronic states increases as compared to the LVO film. Moreover, due to the presence of Ni 3$d$-O 2$p$ hybridized bands near the E$_F$, the energy required to transfer an electron from occupied O 2$p$ state to unoccupied V 3$d$ states decreases compared to LVO film as shown in the Fig. 7(b). Such modulation in energy separation between different states with Ni doping in LVO would arise due to the modification in lattice parameter as observed in the crystal structure analysis. Although, sufficient doping of Ni in LVO would lead to the emergence of delocalized coherent states at E$_F$, the existence of an intense incoherent states in LaNi$_x$V$_{1-x}$O$_3$ films suggests an overwhelming presence of correlation effects nearly localizing charge carriers. The schematics of the overall band structure as derived from these experimental observations are illustrated in Fig. 7(c), which exhibits a tale of evolution of bands from LaVO$_3$ to LaNiO$_3$ as Ni is doped in the system.

**Discussions:**

Due to the smaller value of $U$ compared to $\Delta$, the early transition metal oxide (TMO) compounds are classified in the Mott-Hubbard regime in contrast to the late TMO compounds which mainly fall in the charge transfer regime as the $U > \Delta$ [30]. Thus, the low energy charge fluctuation in Mott-Hubbard insulator is of the type $d_i^n d_j^n \rightarrow d_i^{n-1} d_j^{n+1}$ while, the same for the charge transfer insulator is $d_i^n d_j^n \rightarrow d_i^n d_j^{n+1}\underline{L}$. Moreover, in charge transfer insulator, if the $\Delta$ is small as compared to the metal ligand hybridization, then the weight of $d^{n+1}\underline{L}$ configuration becomes dominant in the ground state and can be classified as a negative $\Delta$ compound [5]. Rare-earth nickelates (RNiO$_3$) compounds are suggested to be placed in the negative $\Delta$ regime [19, 62]. Though, the insulating and metallic nature of negative Δ materials is strongly dependent on the strength of inter-cluster hybridization, resultantly RNiO$_3$ shows MIT, except LNO [14,64]. The doping of late-TM element Ni in Mott-Hubbard insulator LVO does not lead to the carrier doping in LaNi$_x$V$_{1-x}$O$_3$ as observed in the electronic structure analysis, rather the changes in ionic radius and consequent modification in crystal structure inherently lead to the modulation in bandwidth of TM-O 2$p$ hybridized band. Generally, both the bandwidth controlled and carrier concentration controlled MIT are governed by the altering of electron correlation effect ($U/W$). However, the doping of Ni in LVO leads to a transition from the Mott-Hubbard insulating regime to negative Δ metallic regime, which has been controlled by the interplay between the $U$ and Δ as shown schematically in Fig. 8. Therefore, unlike in Mott-Hubbard insulators where the strong Coulomb interaction between the $d$ electrons decides the electronic structure and physical properties of the system, the properties of Ni doped LVO films are governed by CT energy, which can provide an additional degree of freedom to optimize material properties. Moreover, the augmentation of Ni 3$d$ spectral DOS in the Mott-Hubbard gap of insulating LVO with Ni doping offers an alternate pathway to study the interplay between the localized and itinerant state of electron and tune the physical properties in a controlled experiment.

**Conclusion:**

The electronic structure of LaNi$_x$V$_{1-x}$O$_3$ thin films grown on LaAlO$_3$ substrates have been analyzed employing the photoemission and absorption spectra at RT. The out-of-plane lattice parameter of grown films decreases with the Ni doping in LVO. Such Ni doping in Mott-Hubbard insulator LVO does not lead to the carrier doping in LaNi$_x$V$_{1-x}$O$_3$ films, rather the modification in crystal structure leads to the modulation in bandwidth of TM 3$d$-O 2$p$ hybridized band near E$_F$. The shift of Ni 3$d$-O 2$p$ hybridized band towards higher energy in LaNi$_x$V$_{1-x}$O$_3$ films endorses the modification of ligand to metal charge transfer energy. In sufficiently doped LaNi$_x$V$_{1-x}$O$_3$ films, besides the presence of incoherent structure, corresponding to the V 3$d$ band essentially localized due to the electron correlation, a coherent structure also appears at the E$_F$ corresponding to the hybridized Ni 3$d$- O 2$p$ band which provides delocalized electrons for conduction. Our present work offers a spectroscopic realization of Mott-Hubbard insulator to negative Δ metal transition at RT.

**Acknowledgements:**

Authors acknowledge Avinash Wadikar, Sharad Karwal, and Rakesh Sah, for helping in PES and XAS measurements.

**Figure captions:**

**Fig. 1:** (a) θ-2θ x-ray diffraction patterns of LaNi$_x$V$_{1-x}$O$_3$ thin films on grown LaAlO$_3$ substrates. (b) Variation of calculated pseudocubic out-of-plane lattice parameters with Ni doping in LVO films, along with expected pseudocubic bulk lattice parameters.

**Fig. 2:** (a) Experimental V 2$p$ and O 1$s$ x-ray photoemission spectra (XPS) of LaNi$_x$V$_{1-x}$O$_3$ ($x$ = 0.0, 0.5) thin films and simulated XPS of V$^{3+}$ state at RT. (b) Zoomed view of the V 2$p_{3/2}$ peak of LaNi$_x$V$_{1-x}$O$_3$ ($x$ = 0.0, 0.5) thin films. (c) Schematic illustration of the emergence and absence of coherent state in metal and insulator. (d) O 1$s$ core level spectra of LaNi$_x$V$_{1-x}$O$_3$ ($x$ = 0.0, 0.5, 1.0) thin films.

**Fig. 3:** (a) RT V 2$p$ near edge x-ray absorption spectra (NEXAS) of LaNi$_x$V$_{1-x}$O$_3$ thin films along with trivalent V$_2$O$_3$ and pentavalent V$_2$O$_5$ compounds. (b) O 1$s$ NEXAS of LaNi$_x$V$_{1-x}$O$_3$ thin films at RT. (c) Experimental and simulated V$^{3+}$ $L$-edge spectra of LaVO$_3$ and LaNi$_{0.8}$V$_{0.2}$O$_3$ thin films at RT.

**Fig. 4:** (a) La $M_{5,4}$ and Ni $L_3$-edge spectra of LaNi$_x$V$_{1-x}$O$_3$ thin films at 300 K. (b) For the Ni $L$-edge spectrum of LNO films, the La 3$d_{3/2}$ absorption peak (indicated by dashed line) was subtracted from raw data. (c) Extracted Ni $L_{3,2}$ edge spectra of LaNi$_x$V$_{1-x}$O$_3$ films (For extraction of Ni $L$-edge of Ni doped LVO films same procedure has been followed as mention in Fig (b)).

**Fig. 5:** (a) RT valence band spectra (VBS) of LaNi$_x$V$_{1-x}$O$_3$ thin films recorded at 56 eV incident photon energy. (b) Zoomed view of VBS of LaVO$_3$ and LaNi$_{0.5}$V$_{0.5}$O$_3$ thin films near E$_F$. (c) Schematic representation of V 3$d$, Ni 3$d$ and O 2$p$ bands in valence band obtained from the VBS of LaNi$_x$V$_{1-x}$O$_3$ thin films.

**Fig. 6:** (a) Extraction procedure of V 3$d$ spectral DOS of LVO film at RT. The dotted line labeled as "BG" represents the rough estimation of inelastic background. See the text for details. (b) Extraction procedure of V 3$d$ spectral DOS of LaNi$_{0.5}$V$_{0.5}$O$_3$ film at RT. (c) Comparison of the extracted V 3$d$ spectral DOS along with calculated V 3$d$ PDOS of strained orthorhombic LVO. (d) Extraction procedure of Ni 3$d$ spectral DOS of LNO film at RT. Inset shows the zoomed view of the difference spectra near E$_F$.

**Fig. 7:** (a) Combined valence and conduction bands of LaNi$_x$V$_{1-x}$O$_3$ thin films at RT. (Inset) zoomed view of the band edge portion of combined spectra of LVO film (b) Schematic illustration of metal V and Ni 3$d$ bands and oxygen 2$p$ bands along with corresponding charge transfer ($\Delta$) and Coulomb correlation ($U$) energies. (c) The schematics of the overall band structure as derived from these experimental observations.

**Fig. 8:** Schematic representation of the transition from Mott-Hubbard insulating regime to negative $\Delta$ metallic regime with Ni doping in LVO film in $U/t$ and $\Delta/t$ phase diagram.

**Figure 1:**

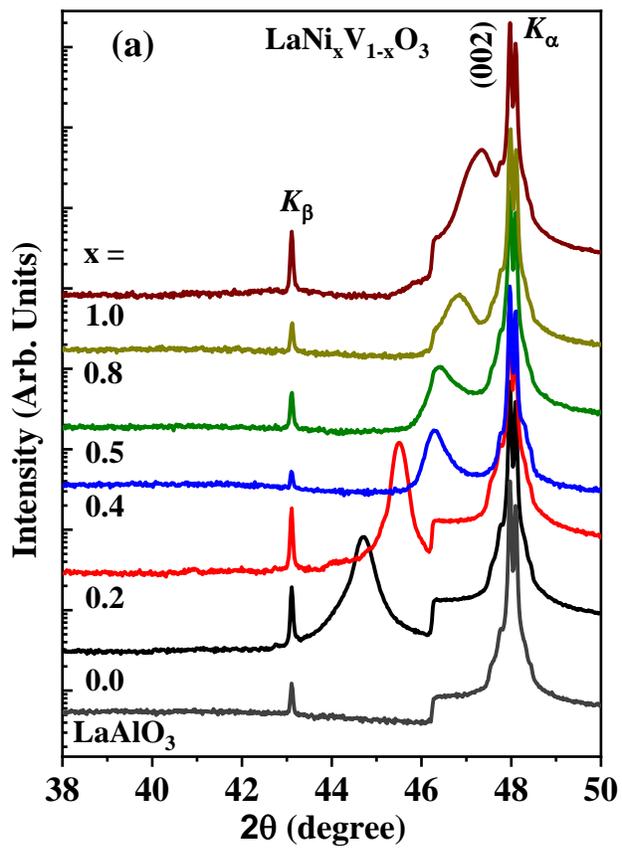

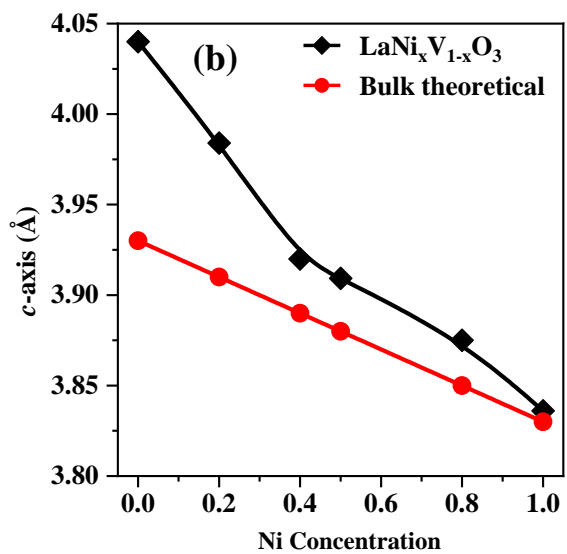

**Figure 2:**

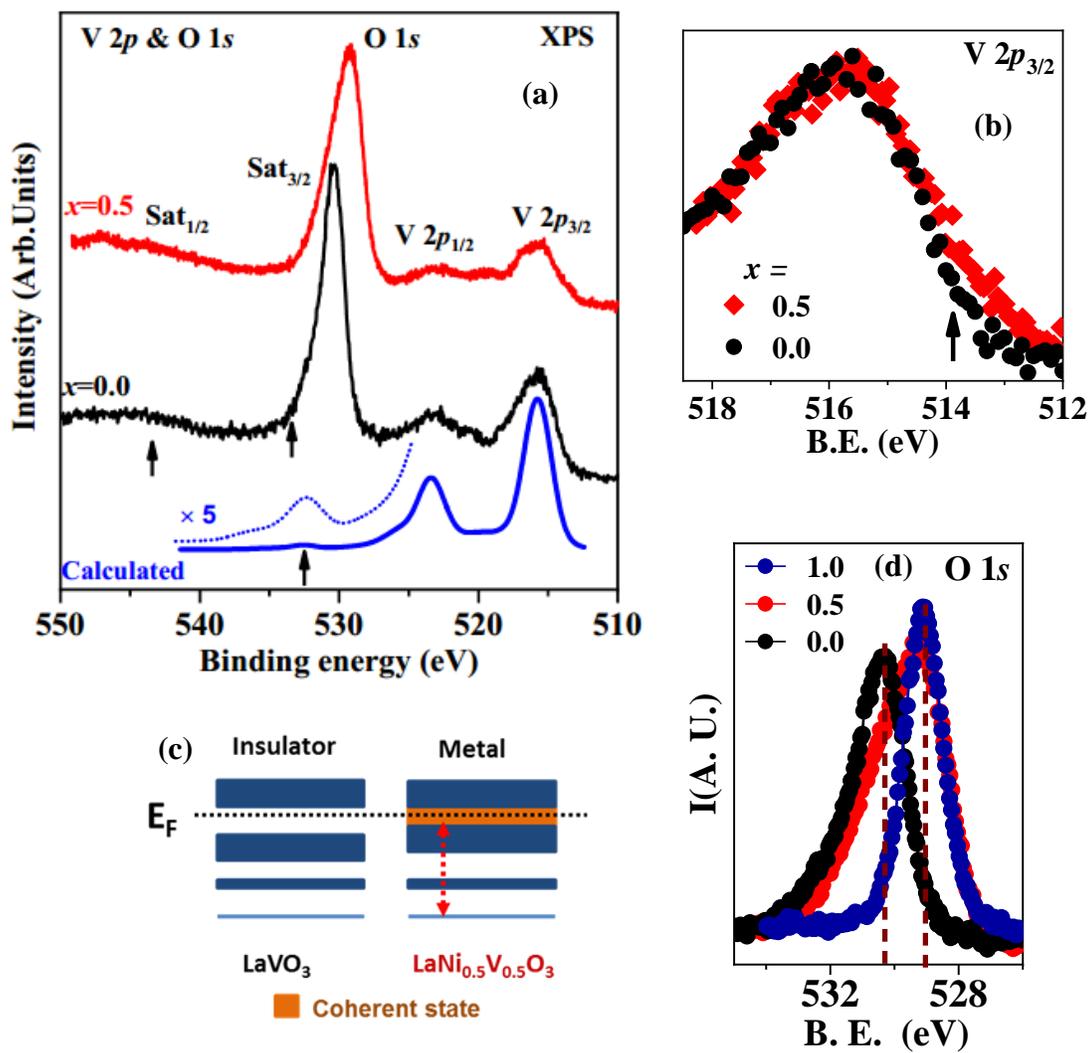

**Figure 3:**

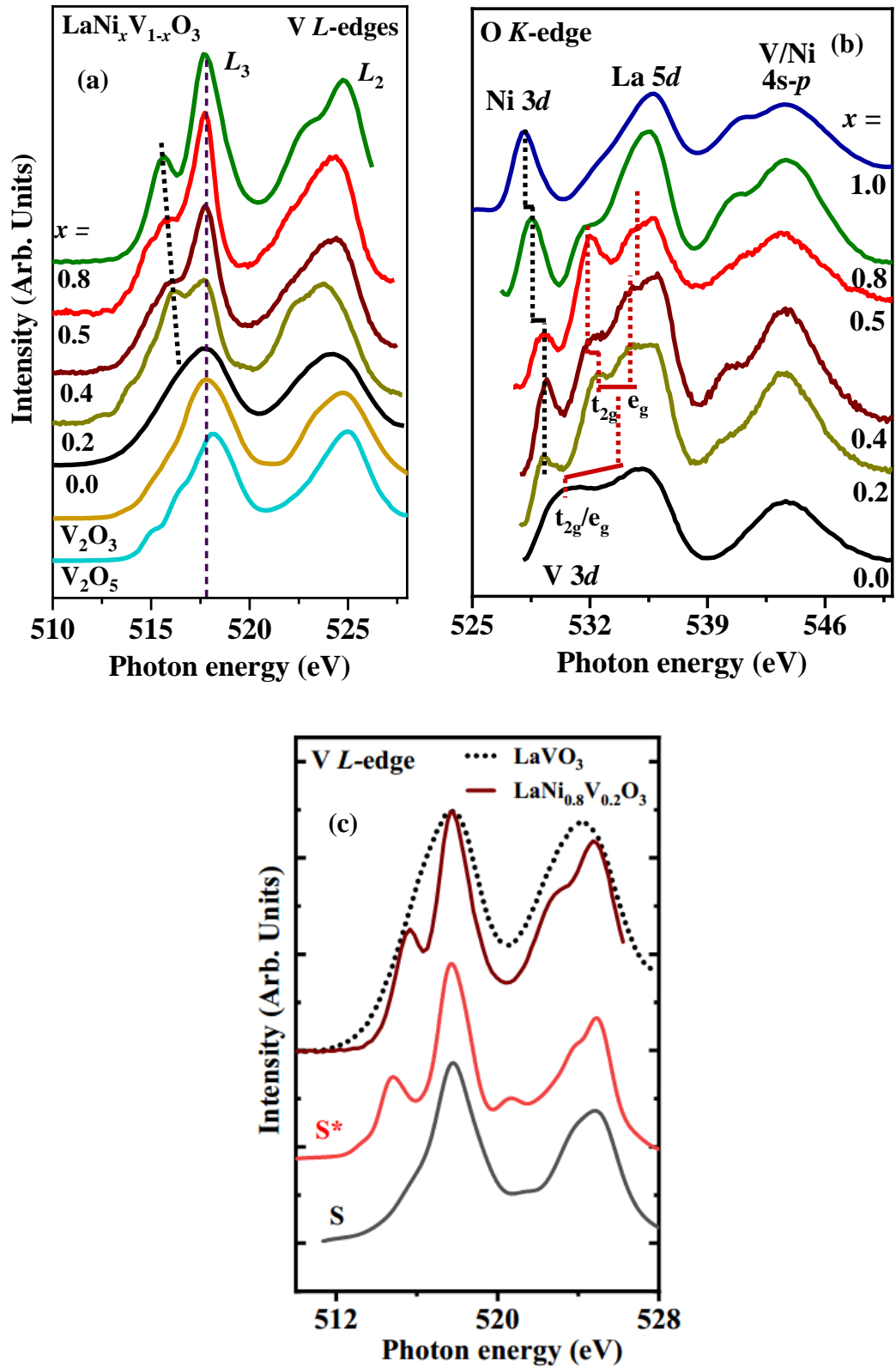

**Figure 4:**

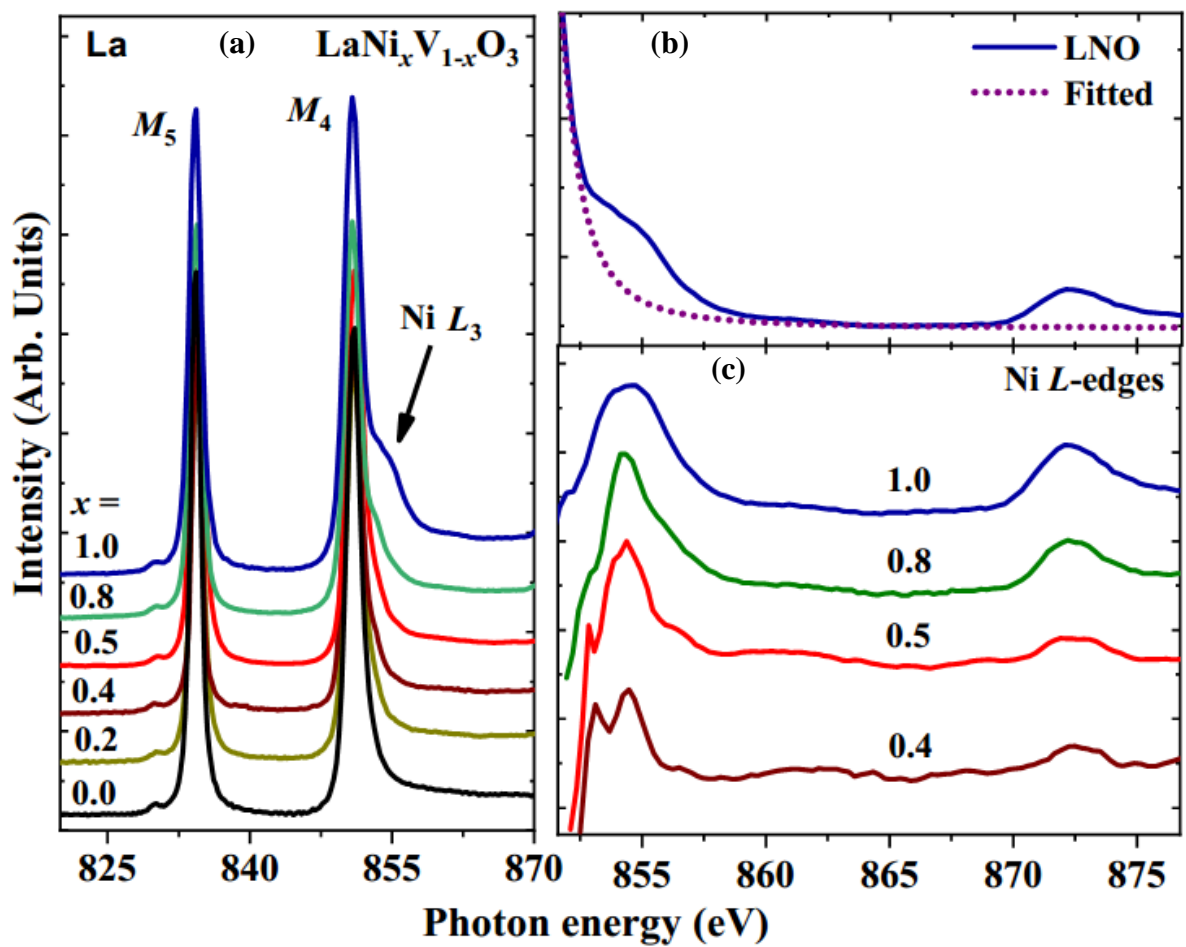

**Figure 5:**

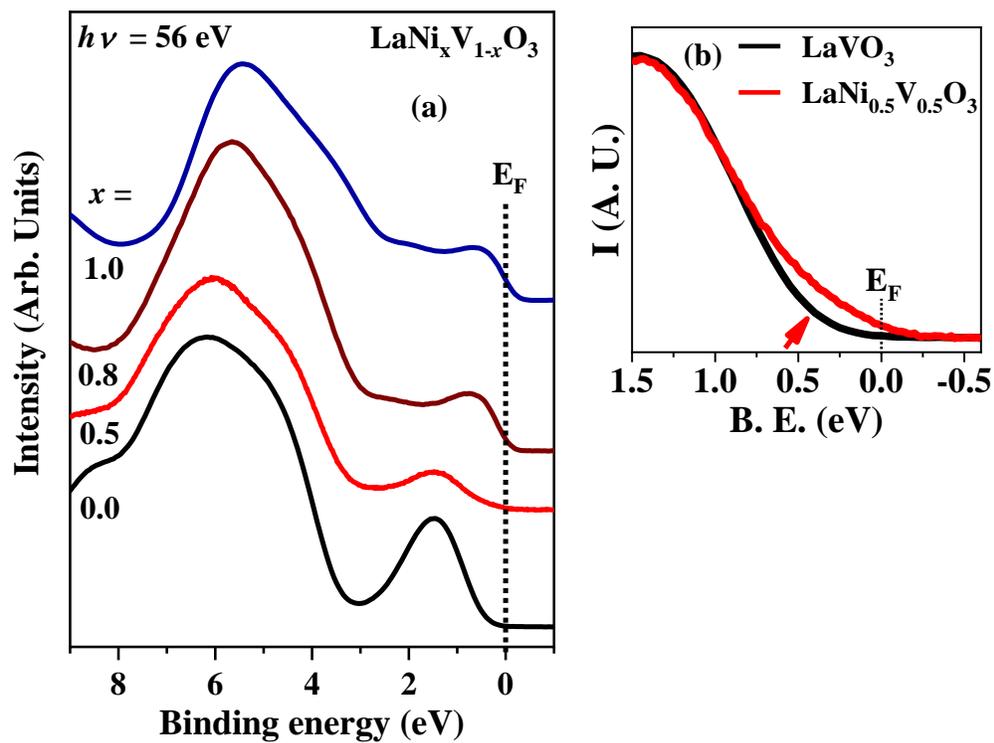

**Figure 6:**

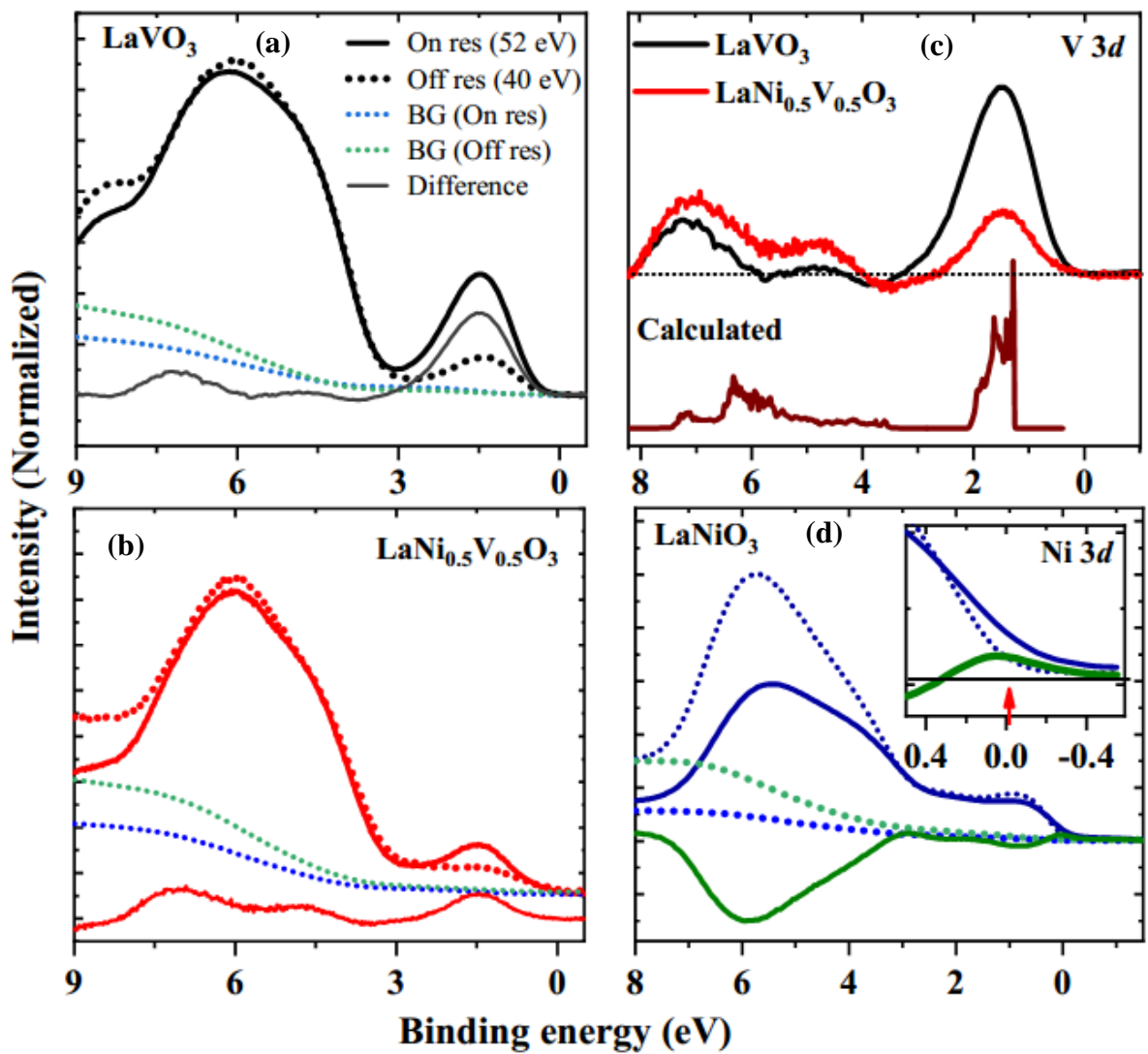

**Figure 7:**

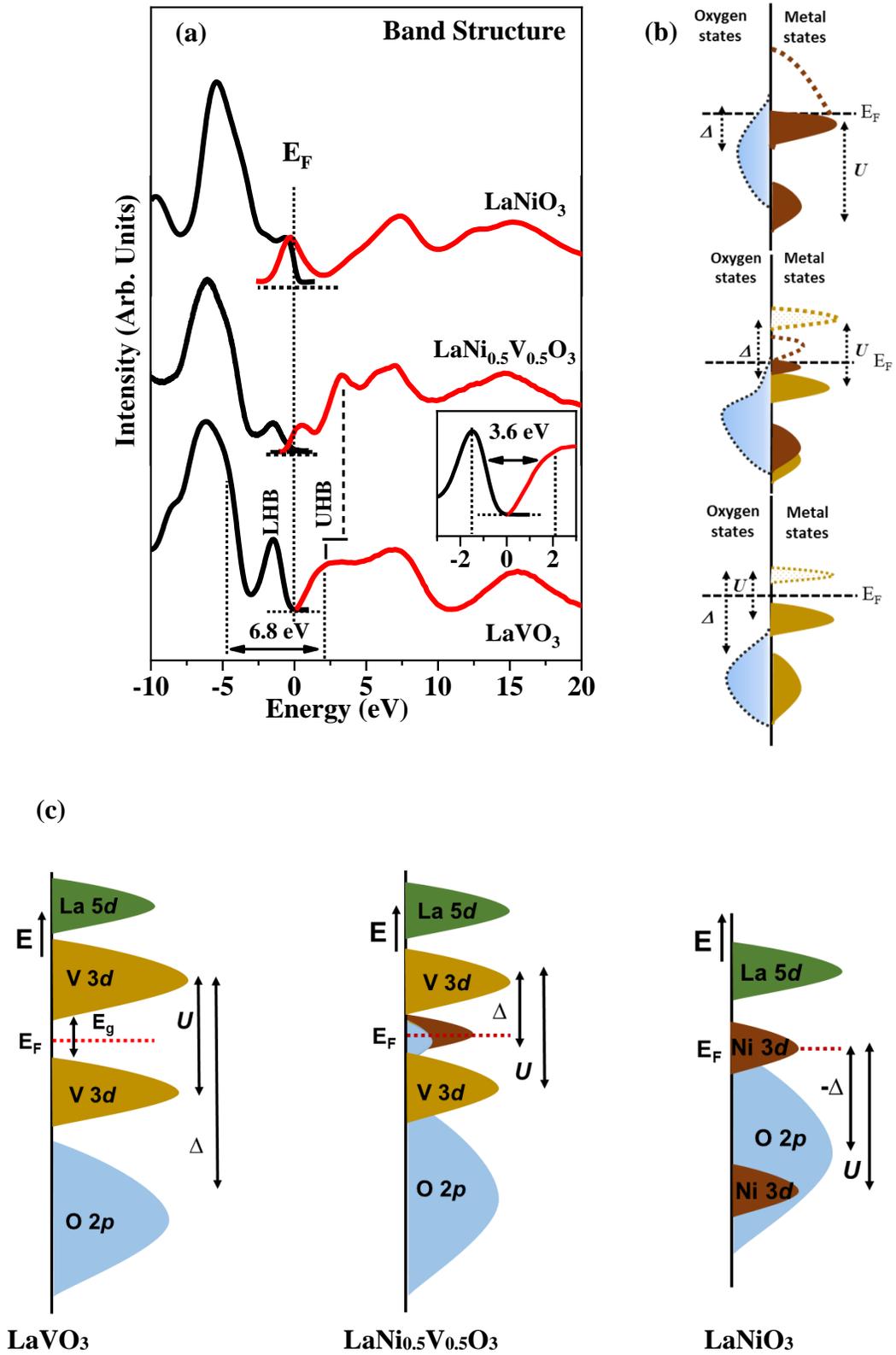

**Figure 8:**

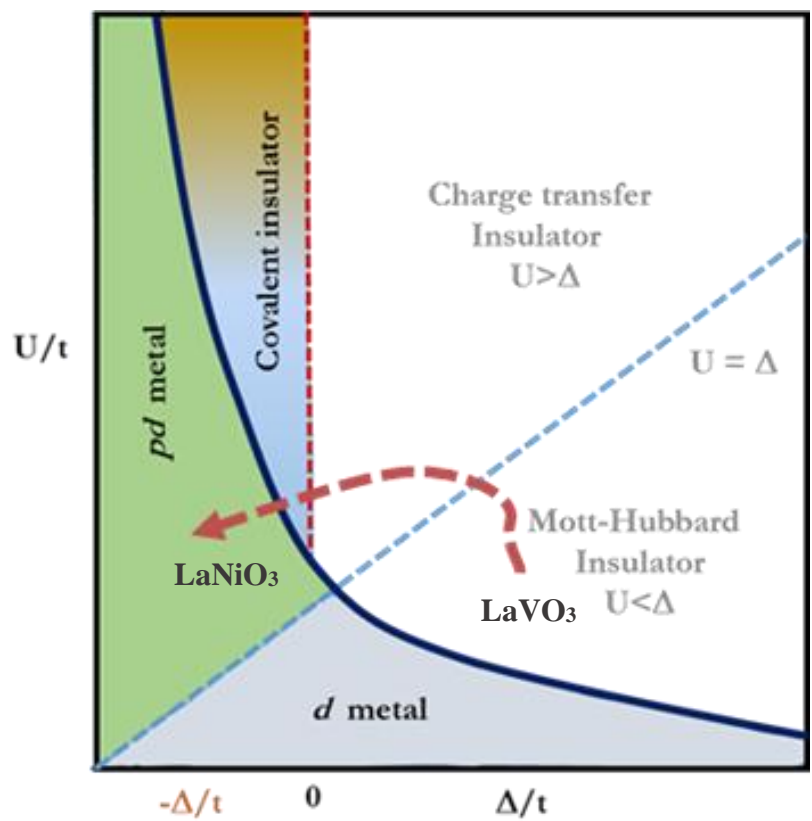